\documentclass[preprint,aps,prd,superscriptaddress,showpacs,showkeys,nofootinbib,notitlepage ]{revtex4-1}
\usepackage{graphicx}
\usepackage{amsmath}
\usepackage{longtable}
\usepackage{color}

\newcommand{\be}{\begin{equation}}
\newcommand{\ee}{\end{equation}}
\newcommand{\bea}{\begin{eqnarray}}
\newcommand{\eea}{\end{eqnarray}}
\newcommand{\etal}{et al.}

\begin{document}

\bibliographystyle{apsrev}

\title{An integrated Sachs-Wolfe effect\,vs\,redshift test for the cosmological parameters}

\author{R. Kantowski}
\email{kantowski@ou.edu}
\affiliation{Homer L.~Dodge Department~of  Physics and Astronomy, University of
Oklahoma, 440 West Brooks,  Norman, OK 73019, USA}

\author{B. Chen}
\email{bchen3@fsu.edu}
\affiliation{Research Computing Center, Department of Scientific Computing,
Florida State University, Tallahassee, FL 32306, USA}
\author{X. Dai}
\email{xdai@ou.edu}
\affiliation{Homer L.~Dodge Department~of  Physics and Astronomy, University of
Oklahoma, 440 West Brooks,   Norman, OK 73019, USA}
\date{\today}
\begin{abstract}
We describe a method using the integrated Sachs-Wolfe (ISW) effect caused by individual inhomogeneities to determine the cosmological parameters, $H_0$,  $\Omega_{\rm m}$, and $\Omega_\Lambda$, etc.
This ISW-redshift test requires detailed knowledge  of the internal kinematics of a  set of individual density perturbations, e.g., galaxy clusters and/or cosmic voids, in particular their density and velocity profiles, and their mass accretion rates.
It assumes the density perturbations are  isolated and imbedded (equivalently compensated) and makes use of the newly found relation between the ISW temperature perturbation of the CMB and the Fermat potential of the lens.
Given measurements of the amplitudes of the temperature variations in the CMB caused by such clusters or voids at various redshifts and estimates of their angular sizes or masses, one can constrain the cosmological parameters.
More realistically, the converse is more likely, i.e., if the  background cosmology is sufficiently constrained, measurement of  ISW profiles of  clusters and voids (e.g., hot and cold spots and rings) can constrain dynamical properties of the dark matter, including accretion, associated with such lenses  and thus constrain the evolution of these objects with redshift.
\end{abstract}



\maketitle

\section{Introduction}
The late time integrated Sachs-Wolfe (ISW) effect \cite{Sachs67}, also called the Rees-Sciama (RS) effect \cite{Rees68}, has recently been suggested (as well and disputed) as the source of observed hot and cold spots in the CMB temperature around some known large scale structures---galaxy clusters and cosmic voids
\cite{Granett08,Granett08b,Planck14}.
By modeling cluster and void density profiles, and by adjusting cluster masses and void depths, observed temperature excesses/deficits can be matched by ISW predictions \cite{Inoue06,Rudnick07,Nadathur12,Hernandez10,Ilic13,Cai14}.
Several proposals also exist to use   lensing of the CMB to determine properties of these clusters  and voids as well as the cosmological parameters \cite{Lavaux12,Melin14,Chantavat14,Hamaus14}.
What we present in this paper is not unrelated to these proposals but offers an easier and more direct method for relating the ISW temperature shifts to the cluster/void structure and the background cosmology.
The conventional approach to determine the ISW effect is to first construct  the ``lensing potential" of a cluster or void from a model of its density profile and then compute the potential's effect on the observed CMB's temperature.
Our approach uses another lensing quantity, the ``Fermat potential" or equivalently the potential part of the time delay, to relate the lens and cosmology to the ISW temperature fluctuations.
Our method of evaluating the ISW effect is  directly related to the lens' mass profile and is more transparent than the conventional approach.
It is simpler to use and requires the construction of only one single function, the potential part of the time delay \cite{Cooke75}.
It is also flexible to use, i.e., the lens structure and/or its evolution can easily be varied and the effects of either are separately discerned.

\begin{figure*}
\includegraphics[width=0.9\textwidth,height=0.18\textheight]{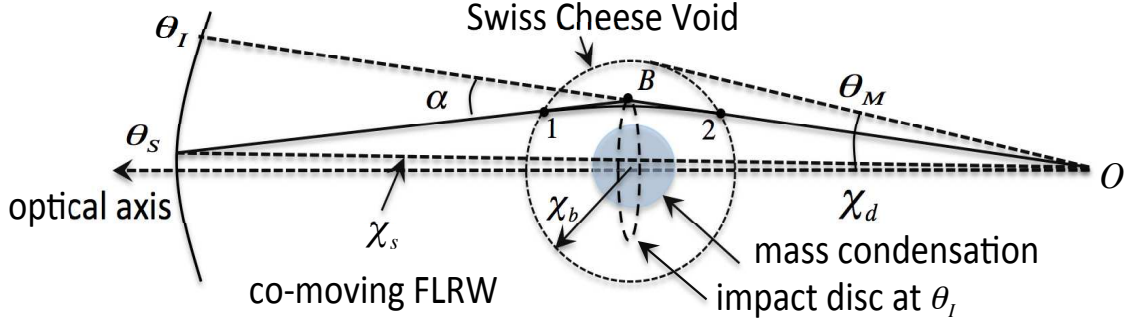}
\caption{ The comoving geometry of an embedded lens at redshift $1+z_d=R_0/R(t_d)$.
Angles $\theta_S$ and $\theta_I$ respectively, are  source and image angles;
$\chi_d$ and $\chi_s$ are the comoving angular distances of the lens and the source.
The (constant) angular size of the void, in lowest order lensing theory,  is $\theta_M\equiv \chi_b/\chi_d$ where $\chi_b$ is the comoving radius of the Swiss cheese void.
	The physical radius of the deflecting lens depends on the cosmic time $t_d,$ i.e.,  $r_d=R(t_d)\chi_b$.
	The shadowed area represents an embedded cluster.
	The dashed circle shows the impact disc of angular radius $\theta_I$, used to compute the included projected mass fraction $f(x)$ of the lens, see Eq.\,(\ref{T}).
         The equivalent figure for a void lens has a mass condensation surrounding a low density central region and a repulsive instead of attractive deflection angle $\alpha$.
}
\label{fig:cheese}
\end{figure*}

We have recently developed the embedded lens theory \cite{Kantowski10,Chen10,Chen11,Kantowski12,Kantowski13,Chen13a,Chen13b} which could be called the Swiss cheese lens theory, or at lowest order, the compensated lens theory.
The theory originated from the Swiss cheese models of general relativity (GR) \cite{Einstein45,Schucking54,Kantowski69}, therefore one can be confident of its gravitational predictions, if GR is indeed the correct theory.
An embedded lens  at redshift  $z_d$ is constructed by first removing a comoving sphere of radius $\chi_b$ from a homogeneous Friedman-Lema\^itre-Robertson-Walker (FLRW) cosmology producing a Swiss cheese void, see Fig.\,\ref{fig:cheese}.
The void has a physical radius $r_d=\chi_b R(t_d)$ at cosmic time $t_d$ that expands with the radius of the background cosmology $R(t)$  but has a  constant angular radius $\theta_M$ as seen by an observer, as the observer ages. In the lowest order lensing theory \cite{Chen11} these radii are related (by embedding) to the Schwarzschild radius $r_{\rm s}$ of the removed mass $M_d$ by
\be
\theta_M =\frac{r_d}{D_d}=\frac{1}{1+z_d}\frac{1}{D_d}\left(\frac{r_{\rm s}}{\Omega_{\rm m}}\frac{c^2}{H_0^2}\right)^{1/3},
\label{thetaM}
\ee
where $D_d$ is the angular diameter distance of the void's center in the background cosmology (e.g., a standard FLRW cosmology), $H_0$ the Hubble constant, and $\Omega_{\rm m}$ the matter density parameter.
The total mass removed $M_d$ is next replaced with any appropriate spherical density while keeping Einstein's equations satisfied throughout the Swiss cheese void and on its time-evolving boundary. The logic for  embedding is simple, by computing the mean density inside larger and  larger spheres centered on a density perturbation, a radius will be reached beyond which the mean density coincides with the FLRW background.
The minimum radius at which this takes place can be chosen as $r_d$.
The simplest such exact Swiss cheese models are constructed by filling the void with one of the Lema\^itre-Tolman-Bondi (LTB) models \cite{Lemaitre33,Tolman34,Bondi47}.
Since we are only interested in the lowest order lensing theory, any non-relativistic fluid whose net mass is the same as the removed Swiss cheese void's mass $M_d$ will suffice. Consequently, models of physical voids must be surrounded by higher density regions and cluster models surrounded by lower density regions.
Such linearized gravitational models are often referred to as compensated \cite{Nottale84,Martinez90,Panek92,Seljak96b,Sakai08,Valkenburg09}.

For spherical density perturbations we have shown in \cite{Kantowski13,Chen13a} that to lowest order an embedded lens can be completely described by its Fermat potential  (equivalent to the sum of the geometrical and potential time delays, $cT=cT_g+T_p$)
 \bea
cT(\theta_S,\theta_I)&=& (1+z_d)\frac{D_dD_s}{D_{ds}}\Bigg[\frac{(\theta_S-\theta_I)^2}{2}
 +\theta_E^2\int_{x}^{1}\frac{f(x',z_d)-f_{\rm RW}(x')}{x'}dx'\Bigg].
\label{T}
\eea
Here $x\equiv\theta_I/\theta_M$ is the normalized image angle, $f(x)\equiv M_{\rm disc}(\theta_I)/M_{\rm disc}(\theta_M)$  is the fraction of the embedded lens' mass projected within the impact disc of angular radius $\theta_I$, and $f_{\rm RW}(x)=1-(1-x^2)^{3/2}$ is the corresponding quantity for the removed co-moving FLRW dust sphere.
At (and beyond) the boundary of the embedded lens, $f(x)=f_{\rm RW}(x)=1$. The angle
$\theta_E=\sqrt{2r_{\rm s}D_{ds}/D_dD_s}$ is the usual Einstein ring angle. Distances
$D_s$ and $D_{ds}$ are  angular diameter distances to the source measured from the observer and the deflector, respectively.
The geometrical part of the time delay $T_g$, i.e., the first term in Eq.\,(\ref{T}), has a universal form whereas the potential part  $T_p$ depends on the individual lens structure.
To construct the Fermat potential all that is needed is a mass density profile $\rho(r,z_d)$ for which
\be\label{Tp}
cT_p(\theta_I,z_d)= 2(1+z_d)r_{\rm s}\int_x^1{\frac{f(x',z_d)-f_{\rm RW}(x')}{x'}{dx'}},
\ee
can be integrated.
All embedded lens properties can be constructed once the specific $T_p(\theta_I,z_d)$ is known.
For example the specific lens equation is given by a $\theta_I$-variation $\delta T(\theta_S,\theta_I)/\delta\theta_I=0.$
In \citep{Chen13a} we have shown that the ISW effect \cite{Sachs67,Rees68} is obtained by a $z_d$-derivative  of $T_p$ (or $T$ since $\partial T_g/\partial z_d\equiv 0$)
\be
\frac{\Delta {\cal T}(\theta_I,z_d)}{{\cal T}}= H_d\,\frac{\partial\, T_p(\theta_I,z_d)}{\partial\, z_d}.
\label{calT}
\ee
In this expression $\Delta {\cal T}$ is the change in the CMB's temperature ${\cal T}$ caused by CMB photons passing through an evolving gravitational lens at impact angle $\theta_I$ . The cosmic-time evolution of the lens is replaced by a dependence on the redshift $z_d$ at which it is seen and the Hubble parameter at that redshift is denoted by $H_d=H(z_d)$.
To compute the ISW effect caused by an embedded lens, we need not only the density profile as required by  conventional lens theory \cite{Schneider92} to compute image properties, but we also need the density profile's evolution rate to compute the $z_d$-derivative.
 Because Eq.\,(\ref{calT}) contains only a first  derivative  we do not need to know the lens' history (i.e., the dynamics of its motion), only its density profile and its velocity distribution at lensing time $z_d$.

A somewhat different connection between the ISW effect and lensing, other than the relation of  Eq.\,(\ref{calT}) to  Eq.\,(\ref{T}), has been  noted by prior analytic work. While investigating corrections to the linear ISW effect caused by keeping nonlinear terms in the dark matter momentum density, \cite{Cooray02,Schafer06,Merkel13} found a term in the time derivative of the lensing potential (the integral of which is conventionally used to compute the ISW effect) that depends on the local deflection angle and the local transverse motion of the gravitational lens. This term is reminiscent of  the transverse Doppler effect that produces a dipole signal in the CMB,  i.e., the Birkinshaw-Gull effect \cite{Birkinshaw83,Gurvits86}.  Even though Eqs.\,(\ref{T})--(\ref{calT}) are tailored for spherical lenses whose centers don't move transversely relative to the CMB,  their evolving projected mass densities $(2\pi x)^{-1}df/dx$ are associated with radially directed divergent momentum densities.  That motion would produce local effects on the transiting photons that could be similarity identified if analyzed conventionally using the lensing potential as done by \cite{Cooray02,Schafer06,Merkel13}.

\section{The principle behind the ISW-z test}

Equation (\ref{calT}) gives the fluctuation in the observed CMB temperature as a function of angular position across a given, possibly evolving, density perturbation (a lens) caused by the ISW effect.
From Eq.\,(\ref{calT}) the ISW signal is seen to depend on the lens' redshift $z_d$, its mass $r_{\rm s}$, its projected fractional density profile $f(x,z_d)$
(including its evolution, i.e., its first derivative  with respect to redshift),
as well as the background cosmology.
We construct the new cosmology test using this simple relation.
By splitting Eq.\,(\ref{calT}) into an amplitude term proportional to the product of the lens mass and the Hubble parameter, times a lens structure dependent term ${\cal S}(\theta_I, z_d)$ we have
\be
\frac{\Delta {\cal T}(\theta_I,z_d)}{\cal T}=  2r_{\rm s}\frac{H_d}{c}\times\, {\cal S}(\theta_I, z_d),
\label{newtest-rs}
\ee
where the lens structure dependent term is defined by
\be
{\cal S}(\theta_I, z_d)\equiv \frac{\partial\, }{\partial\, z_d} \left[(1+z_d)\int_x^1{\frac{f(x',z_d)-f_{\rm RW}(x')}{x'}{dx'}}\right].
\label{S}
\ee
If the lens mass and structure are known, the amplitude of $\Delta {\cal T}(\theta_I,z_d)/\cal T$ at the lens' center ($\theta_I=0$) can in principle be used to determine the Hubble parameter $H(z_d)$.
In practice, to apply Eq.\,(\ref{newtest-rs}) to a cold or hot spot associated with a single void or cluster lens it must be averaged over the aperture  of the detector, i.e., $\Delta {\cal T}(\theta_I,z_d)$ and  ${\cal S}(\theta_I, z_d)$ are replaced by  their averaged values, $\Delta {\cal T}(z_d)$ and  ${\cal S}(z_d)$.
If a set of clusters and/or voids can be found whose redshifts, masses, and evolving structures can be determined, then ${\cal S}(\theta_I, z_d)$ and ${\cal S}(z_d)$ can be determined.
Given the CMB temperature data at the positions of these clusters and/or voids, Eq.\,(\ref{newtest-rs}) will determine the Hubble parameter $H_d$ as a function of $z_d$.  The redshift dependent Hubble parameter can then
 be used to constrain all the cosmological parameters.
In Sec.\,\ref{sec:Examples} we illustrate the procedure by applying it to simple top-hat cluster and void models.

The above form of the ISW-z test assumes the mass of the lensing cluster or void is known; however
for cosmic voids, radii can be more easily determined than masses \cite{Sutter12}.
We now present a second form of the ISW-redshift  test preferable for such voids.
This second test requires knowledge  of the energy content of the FLRW background before it can be applied.
We construct this form of ISW-redshift test by looking at the central region of the void or cluster, eliminating $r_{\rm s}$ from Eq.\,(\ref{calT}) by using Eq.\,(\ref{thetaM}), and dividing by the cube of the angular radius of the Swiss cheese void $\theta_M^3$ to obtain the $H_0$ and $r_{\rm s}$ independent result
\be
\frac{\Delta {\cal T}(z_d)/{\cal T}}{(\theta_M)^3 }= {\cal C}(z_d)\times\, {\cal S}(z_d),
\label{newtest-thetaI}
\ee
where the pure curvature dependent part ${\cal C}(z_d)$ is defined by
\be
{\cal C}(z_d)\equiv 2\Omega_{\rm m}\frac{H_d}{H_0}\left[(1+z_d) D_d \frac{H_0}{c}\right]^3,
\label{C}
\ee
and the lens structure dependent term ${\cal S}(z_d)$ is again defined by Eq.\,(\ref{S}).

By replacing the Hubble parameter $H_d$ and the angular diameter distance $D_d$ by functions of the curvature and redshift, assuming for example a simple possibly evolving dark-energy $\Lambda$CDM like gravity source with $p_\Lambda/\rho_\Lambda c^2=[w_0+w_a z/(1+z)]$, we have
\bea
&&H_d/H_0\equiv E(z_d)=\nonumber\\
&&\sqrt{\Omega_\Lambda(1+z_d)^{3(1+w_0+w_a)}e^{-3w_az_d/(1+z_d)}+\Omega_{\rm m}(1+z_d)^3+(1-\Omega_{\rm m}-\Omega_\Lambda)(1+z_d)^2}\,,
\label{E}
\eea
and
\be
(1+z_d)D_d\frac{H_0}{c}=\frac{1}{\sqrt{|1-\Omega_{\rm m}-\Omega_\Lambda|}}{\rm Sinh}\left[\sqrt{|1-\Omega_{\rm m}-\Omega_\Lambda|}\int_0^{z_d} \frac{dz}{E(z)}\right],
\label{Dd}
\ee
where ${\rm Sinh}(x) = \sin(x),$ $x,$ and $\sinh(x)$ for a closed, flat, or open universe, respectively. The  conventional $\Lambda$CDM cosmology is recovered when $w_0=-1$ and $w_a=0$.
The curvature part of Eq.\,(\ref{newtest-thetaI}) becomes
\be
{\cal C}(z_d)\equiv 2\Omega_{\rm m}\,E(z_d)\left\{\frac{1}{\sqrt{|1-\Omega_{\rm m}-\Omega_\Lambda}|}{\rm Sinh}\left[\sqrt{|1-\Omega_{\rm m}-\Omega_\Lambda|}\int_0^{z_d} \frac{dz}{E(z)}\right]\right\}^3.
 \label{curvature}
\ee
The sensitivity of ${\cal C}(z_d)$ to the cosmological parameters can be seen by its series expansion for
$z_d\ll1$
\bea
{\cal C}(z_d)&\approx&  2\Omega_{\rm m}z_d^3
\Biggl[ 1-\frac{1}{2}\left(1+\frac{1}{2}\Omega_{\rm m}+\frac{1}{2}\Omega_\Lambda\,[1+3w_0]\right)z_d\nonumber\\
&+&\frac{1}{4}\left(3-\Omega_{\rm m}-[1-3w_0]\,\Omega_\Lambda +\frac{1}{4}\, \bigl(\Omega_{\rm m}+\Omega_\Lambda[1+3w_0]\bigr)^2 \right)z_d^2
+{\cal O} [z_d^3]
\Biggr].
\label{series}
\eea
It is obviously most sensitive to $\Omega_{\rm m}$ and doesn't even depend on $w_a$ to this order.


\begin{figure*}
\begin{center}$
\begin{array}{cc}
\hspace{-15pt}
\includegraphics[width=0.53\textwidth,height=0.35\textheight]{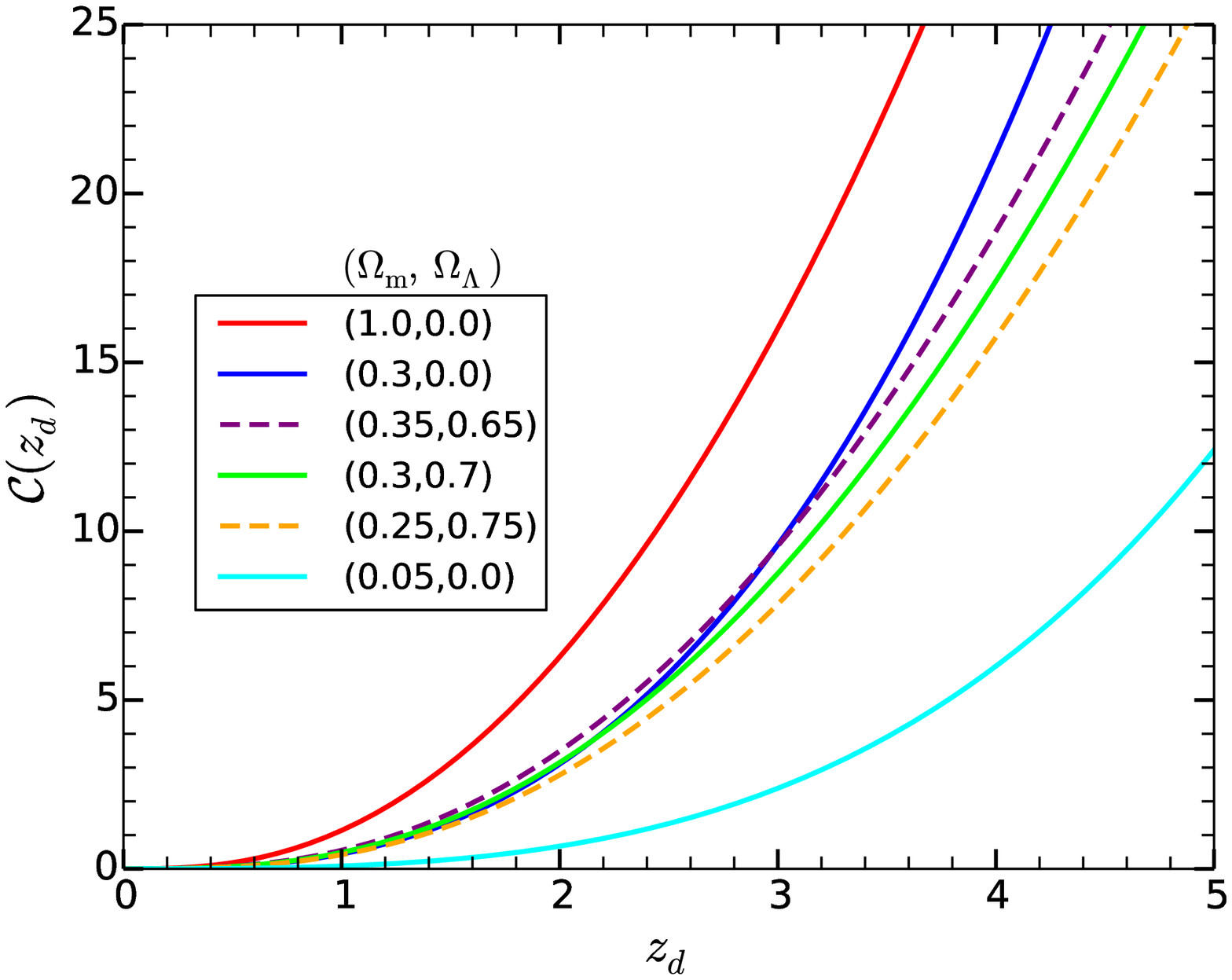}
\hspace{5pt}
\includegraphics[width=0.53\textwidth,height=0.35\textheight]{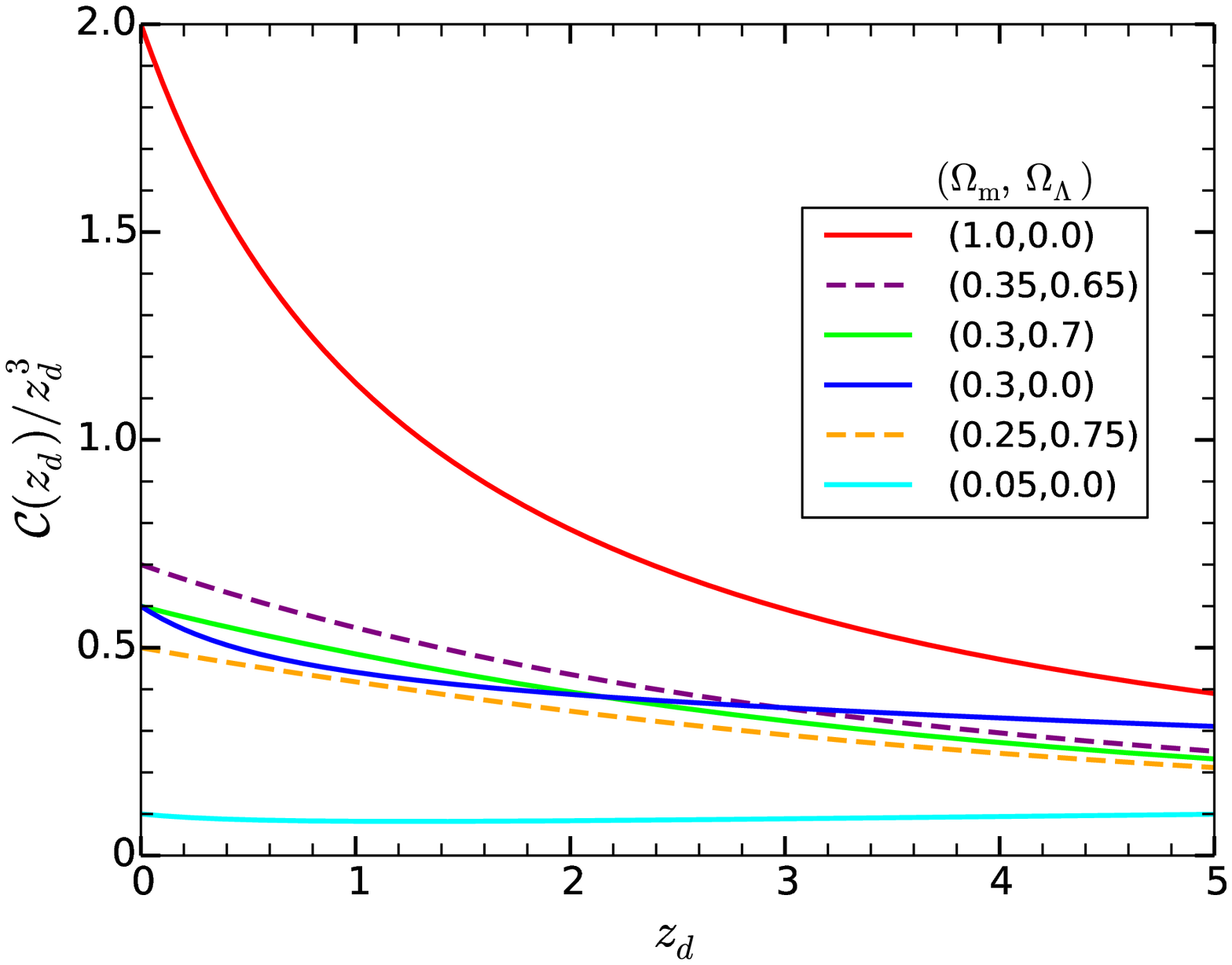}
\end{array}$
\end{center}
\caption{Plots of the curvature term ${\cal C}(z_d)$ from Eq.\,(\ref{curvature}) for four background cosmologies (the solid curves from top to bottom): the Einstein de Sitter universe,  $(\Omega_{\rm m},\Omega_\Lambda)=(1,0),$ red; a dark matter only universe, $(\Omega_{\rm m},\Omega_\Lambda)=(0.3,\,0),$ blue; a  $\Lambda$CDM universe, $(\Omega_{\rm m},\Omega_\Lambda)=(0.3,\,0.7),$ green; and a baryonic matter only universe, $(\Omega_{\rm m},\Omega_\Lambda)=(0.05,\,0),$ cyan. The two dashed curves show the extent to which
${\cal C}(z_d)$ varies for small changes about $(\Omega_{\rm m},\Omega_\Lambda)=(0.3,\,0.7)$. In the right panel the small $z_d$ dependence is factored out.}
\label{fig:C2}
\end{figure*}

We plot the curvature part, Eq.\,(\ref{curvature}), in the left panel of Fig.\,\ref{fig:C2} for four familiar cosmologies as solid curves: the Einstein de Sitter (EdS) universe,  $(\Omega_{\rm m},\Omega_\Lambda)=(1,0)$; a dark matter only universe, $(\Omega_{\rm m},\Omega_\Lambda)=(0.3,\,0)$;  a $\Lambda$CDM universe, $(\Omega_{\rm m},\Omega_\Lambda)=(0.3,\,0.7)$; and a baryonic matter only universe, $(\Omega_{\rm m},\Omega_\Lambda)=(0.05,\,0)$.
We also indicate the range of variation in  ${\cal C}(z_d)$ for flat $\Lambda$CDM models by plotting, as dashed curves, the $(\Omega_{\rm m},\Omega_\Lambda)=(0.35,0.65)$ and $(0.25,0.75)$ cases.
In the right panel of Fig.\,\ref{fig:C2} we plot these same curves but with the common factor $z_d^3$ divided out, i.e., ${\cal C}(z_d)/z_d^3.$
A significant dependence on $\Omega_{\rm m}$ is easily seen.


\begin{figure*}
\begin{center}
\includegraphics[width=0.53\textwidth,height=0.35\textheight]{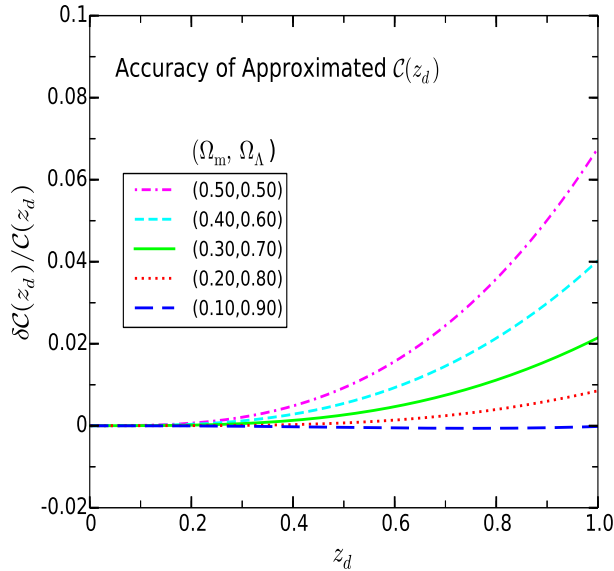}
\end{center}
\caption{The fractional difference of the series approximation for ${\cal C}(z_d)$ as given by Eq.\,(\ref{series}) and the exact value as given by Eq.\,(\ref{curvature}) is plotted for five flat cosmologies  (from top to bottom):  $(\Omega_{\rm m},\Omega_\Lambda)=(0.50,0.50),$ dot-dashed magenta; $(0.40,0.60)$ dashed cyan; $(0.30,0.70)$  solid green; $(0.20,0.80)$ dotted orange; $(0.10,0.90)$ dashed blue.}
\label{fig:series}
\end{figure*}

\begin{figure*}
\begin{center}$
\begin{array}{cc}
\hspace{-15pt}
\includegraphics[width=0.53\textwidth,height=0.35\textheight]{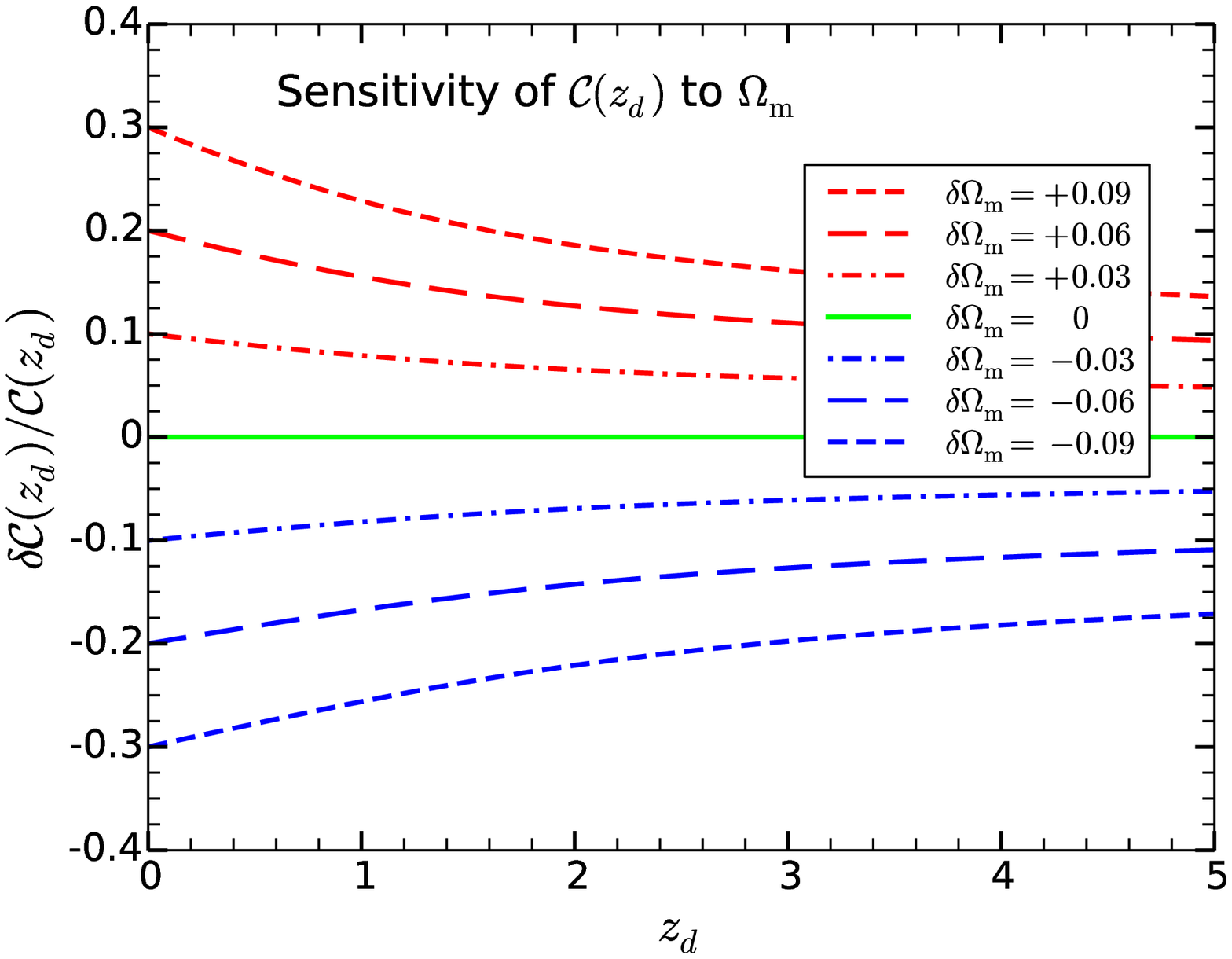}
\hspace{5pt}
\includegraphics[width=0.53\textwidth,height=0.35\textheight]{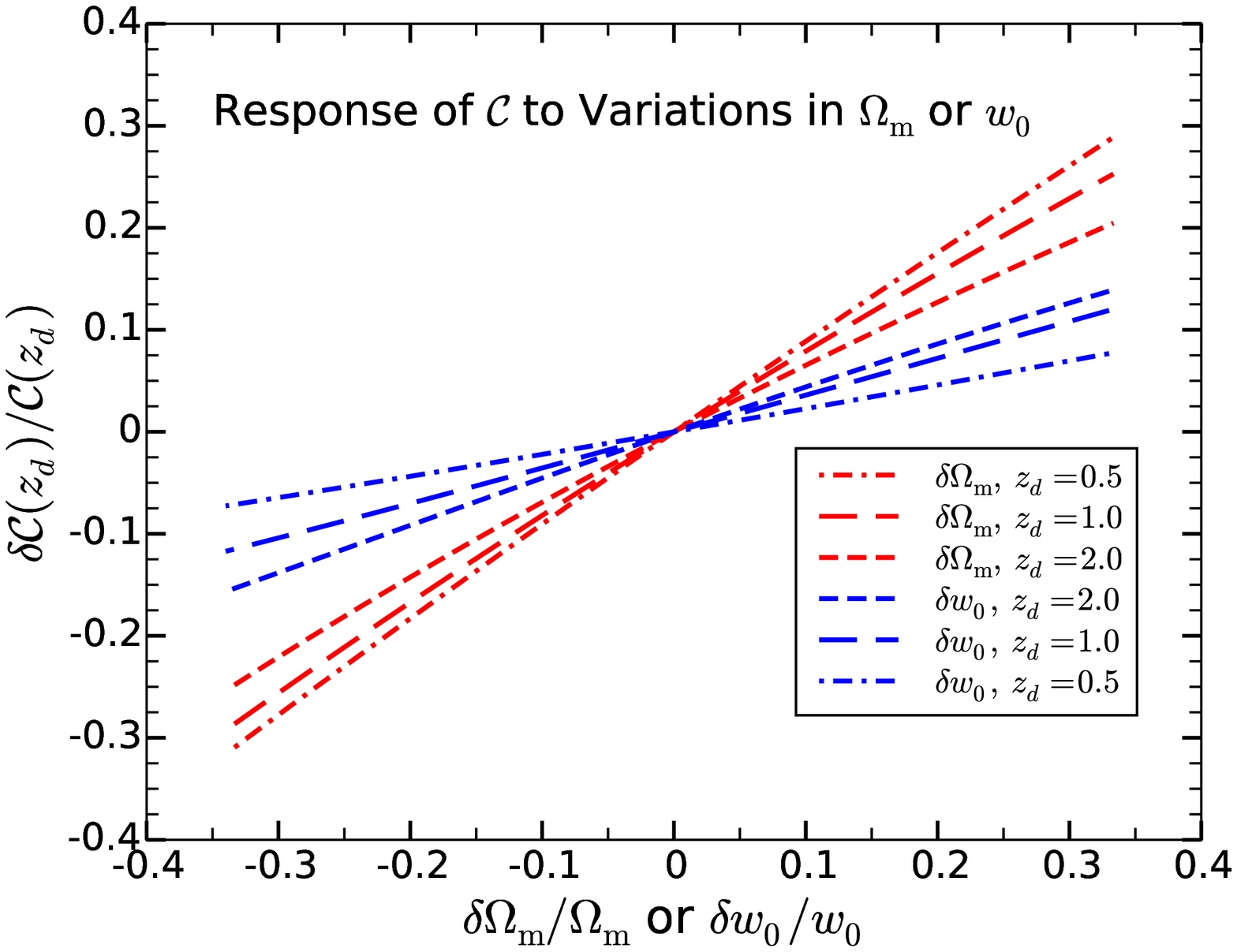}
\end{array}$
\end{center}
\caption{(Left) Sensitivity of ${\cal C}(z_d)$ to the matter density parameters $\Omega_{\rm m}$.
The fractional change in the curvature ${\cal C}(z_d)$ given by Eq.\,(\ref{curvature}) is plotted as a function of the redshift $z_d$ for several \underbar{flat} $\Lambda$CDM cosmologies close to $\Omega_{\rm m}=0.3,$  i.e.,  $\delta{\cal C}/{\cal C}\equiv [{\cal C}(0.3+\delta\Omega_{\rm m})-{\cal C}(0.3)]/{\cal C}(0.3)$ is plotted for $\delta \Omega_{\rm m}=0,\pm0.03,\pm0.06,$ and $\pm0.09.$
(Right) The linear response of the curvature term (relative to the fiducial $\Lambda$CDM cosmology) as a function of variations of $\Omega_{\rm m}$ for lens redshift $z_d=0.5,$ 1.0, and 2.0 (respectively dot-dashed, long dashed, and dashed curves) is plotted as the three red curves  (larger slopes). The three blue curves (smaller slopes) are similarly produced by variations of $w_0$ about $w_0=-1$.
For a lens at redshift $z_d\approx 1,$ a measurement of ${\cal C}$ within 10\% constrains $\Omega_{\rm m}$ and $w_0$ up to about 12\% and 25\%, respectively.}
\label{fig:C4}
\end{figure*}

In Fig.\,\ref{fig:series} we  show the accuracy of the series approximation for ${\cal C}(z_d)$ as given by the first 3 terms of  Eq.\,(\ref{series}) by plotting the fractional difference between ${\cal C}(z_d)$ as given by Eq.\,(\ref{series}) and  ${\cal C}(z_d)$ as given by Eq.\,(\ref{curvature}) for the indicated cosmologies.
For lenses of redshift up to 0.5 the series approximation is accurate to 1\%.
In Fig.\,\ref{fig:C4} we plot, for flat $\Lambda$CDM models, the fractional change in ${\cal C}(z_d)$ for various $(\Omega_{\rm m},\Omega_\Lambda)$ values compared to ${\cal C}(z_d)$  for the (0.3, 0.7) universe.
From the plots in the left panel we see that ${\cal C}(z_d)$  is more sensitive to $\Omega_{\rm m}$ at smaller redshifts,  and from the red curves on the right we see the uncertainty in measurement of the curvature term scales roughly linearly with that in $\Omega_{\rm m}$.
Combining these observations with Fig.\,\ref{fig:series} we see that the series approximation should be accurate enough to determine $\Omega_{\rm m}$ up to about 2\% at redshifts up to $z_d\approx 0.7$ if the Universe is in the neighborhood of  (0.3, 0.7).


\begin{figure*}
\begin{center}
\includegraphics[width=0.53\textwidth,height=0.35\textheight]{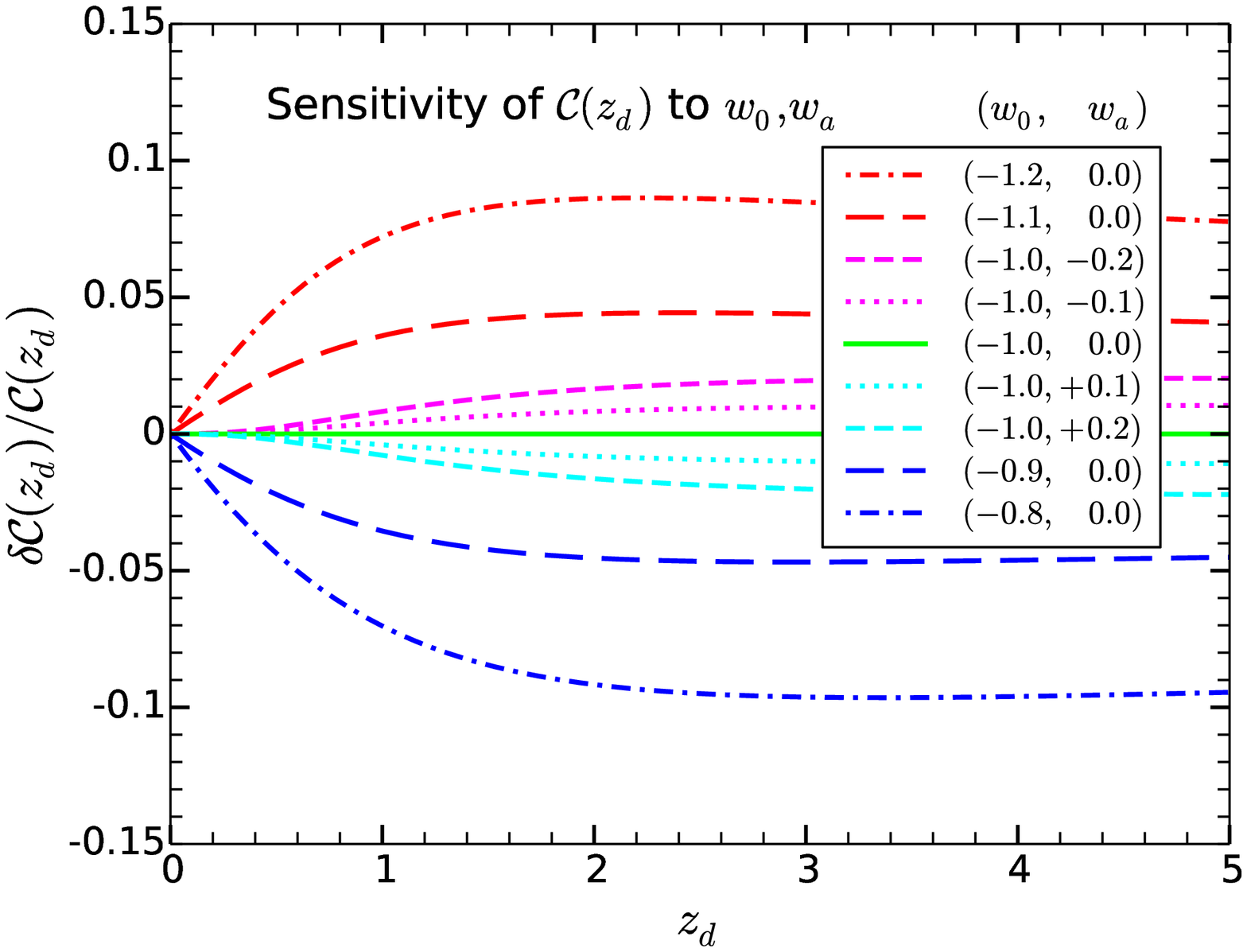}
\end{center}
\caption{Sensitivity of ${\cal C}(z_d)$ to the dark energy parameters $w_0$ and $w_a.$
For all plotted cosmologies  $(\Omega_{\rm m},\Omega_\Lambda)=(0.3,\,0.7)$ but $w_0$ and $w_a$  are allowed to respectively vary about -1 and 0, i.e., we test for $\delta w_0=0,\pm0.1,\pm0.2,$ and $\delta w_a=0,\pm0.1,\pm0.2.$
The curvature term ${\cal C}(z_d)$ is more sensitive to variations in $w_0$ than $w_a$, see Eq.\,(\ref{series}).}
\label{fig:w0wa}
\end{figure*}

The  blue curves plotted in the right panel of Fig.\,\ref{fig:C4} and the curves plotted in  Fig.\,\ref{fig:w0wa} compare curvature functions of  Eq.\,(\ref{curvature}) for the concordance $\Lambda$CDM universe with universes whose $\Lambda$ like field has a dynamical equation of state $p_\Lambda/\rho_\Lambda c^2=[w_0+w_a z/(1+z)]$. We have varied $(w_0,w_a)$ about $(-1, 0)$ keeping $(\Omega_{\rm m},\Omega_\Lambda)=(0.3, 0.7)$ without placing any physical constraint on their values.
For the range of variation of the two parameters shown, the maximum sensitivity is only reached beyond $z_d=1$.


Cosmic voids identified through galaxy surveys generally have low redshifts $z\lesssim 0.5$ \cite{Abazajian09,Sutter13}, and as can be seen in  Eq.\,(\ref{series}) and the right panel of Fig.\,\ref{fig:C2}, at small redshifts the curvature term is dominated by the mass density parameter $\Omega_{\rm m}$. Consequently, this test should place its strongest constraint on  $\Omega_{\rm m}.$

\section{Examples\label{sec:Examples}}

The structure term ${\cal S}(\theta_I,z_d)$ has to be accurately modeled before we can use either of the two tests presented in the previous section to constrain the cosmological parameters.
Modeling strong gravitational lenses  (galaxies or clusters of galaxies) traditionally requires only the density profiles $\rho({\bf r})$ of the lenses, whereas modeling ISW effects requires
the additional knowledge of the first time derivative of  $\rho({\bf r})$ at lensing (equivalent to the first derivative with respect to $z_d$).
Even if we assume that galaxy clusters are virialized, their density contrasts with respect to the FLRW background evolve with redshift, and so do their projected fractional mass profiles $f(x,z_d)$.
To numerically evaluate Fermat potentials for compensated cluster lens models with realistic profiles, e.g., cluster lenses with profiles such as the \underbar{embedded} Navarro-Frenk-White (NFW) profile \cite{NFW96} is straight forward but analytical evaluation is challenging .
Dark matter density profiles for cosmic voids are currently estimated by stacking and averaging galaxy counts over large numbers of voids.
This assumes that luminous matter as tracers of dark matter is not significantly biased and
even if correct,  far less is known about void evolution than about cluster evolution.
There are hints indicating that voids can be deep in the central regions, with $\delta \lesssim -0.8$ near the void center \cite{Sutter12}.
If this is indeed the case, then $\delta$ might be evolving very slowly (already approaching its lower bound of $-1$) and the $z_d$ dependence in $f(x,z_d)$ might be neglegable. If so the structure term would consequently  be easier to model. Voids would simply be expanding with the background and the  ISW effect would be determined by the time-delay contribution alone \cite{Chen13a}.
The ISW-z test might be more fruitfully applied to  cosmic voids than galaxy clusters because it  is very sensitive to cluster accretion (see below) and the CMB temperature seen through a cluster is contaminated by hot gas emissions from the cluster itself and from other secondary anisotropies such as the Sunyaev-Zeldovich (SZ) effect \cite{Sunyaev80,Birkinshaw99}.

\begin{figure}[!htbp]
\begin{center}$
\begin{array}{cc}
\hspace{-10pt}
\includegraphics[width=0.5\textwidth,height=0.3\textheight]{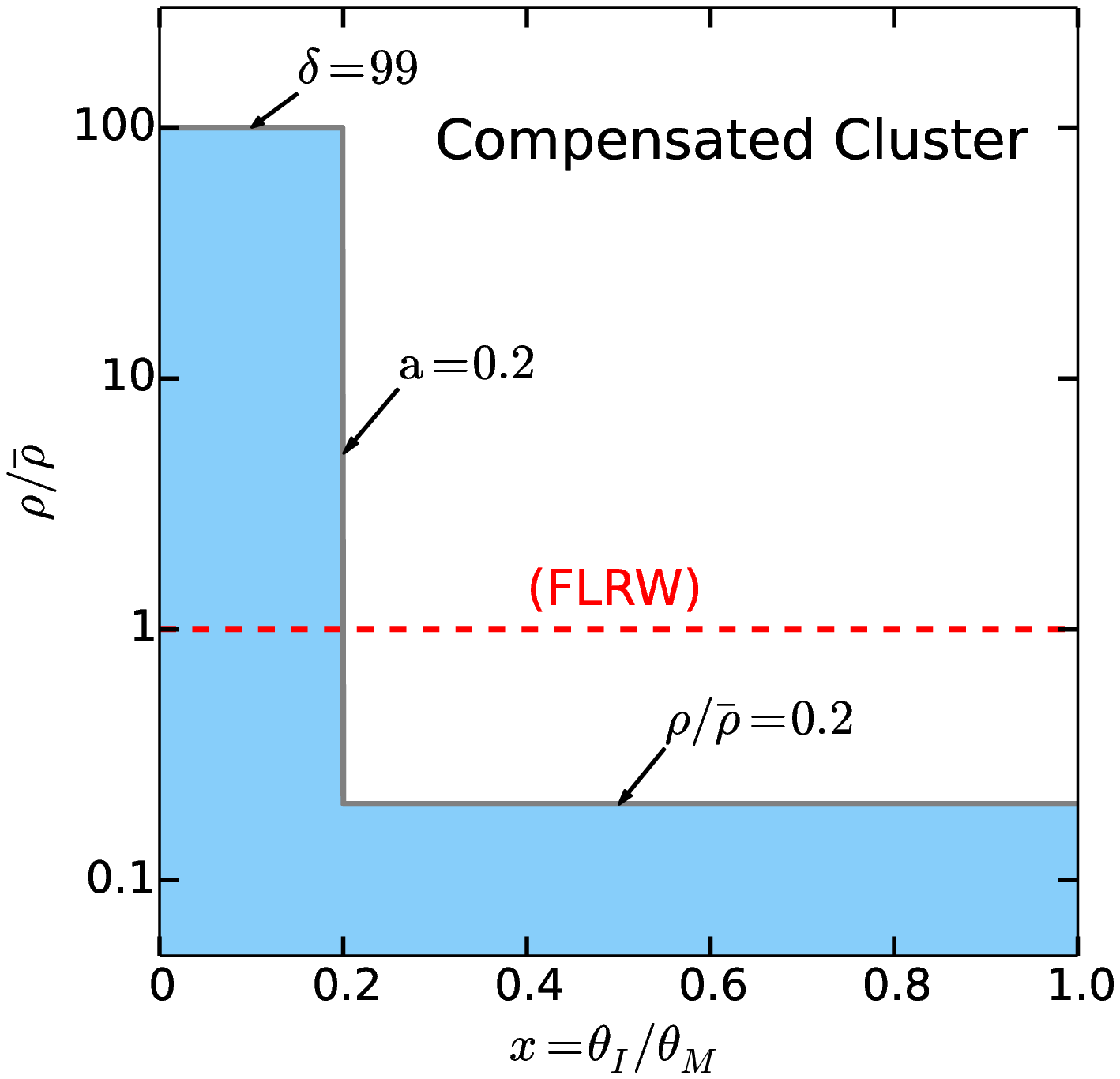}
\hspace{15pt}
\includegraphics[width=0.5\textwidth,height=0.3\textheight]{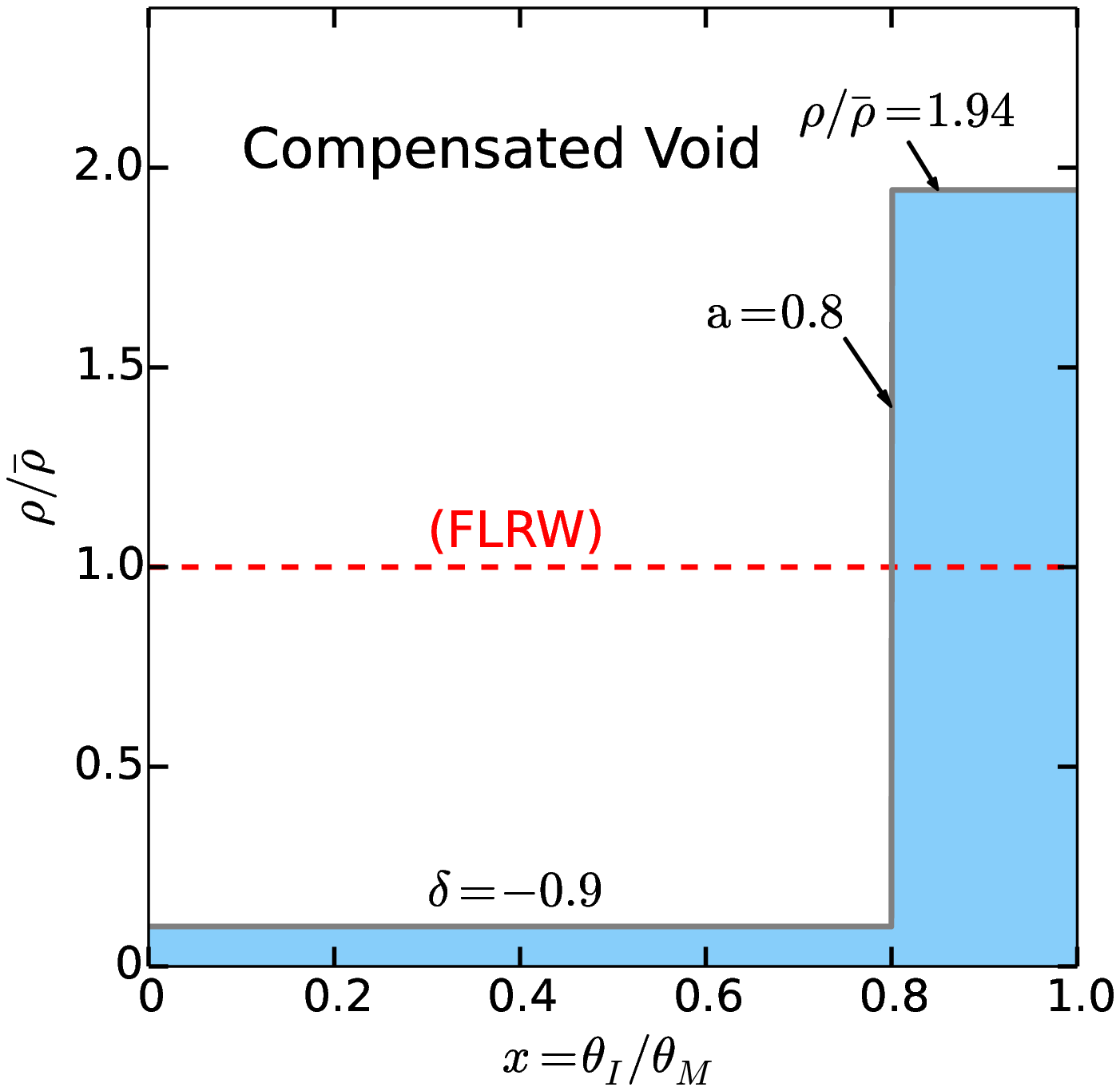}
\end{array}$
\end{center}
\caption{Compensated top-hat models for a cluster on the left ${\mathtt a}=0.2, \delta=99$ and a void on the right ${\mathtt a}=0.8, \delta=-0.9$. The well surrounding the cluster (and the wall surrounding the void)  begins at physical radius $r={\mathtt a}\,r_d$. }
\label{fig:ClusterVoid}
\end{figure}

As a first attempt to illustrate the procedure of constructing the structure term ${\cal S}(z_d)$, we approximate cosmic density perturbations by a two-parameter family of either top-hat models for clusters or inverted top-hat models for voids, see Fig.\,\ref{fig:ClusterVoid}.
Both the cluster and void models are compensated with density profiles defined as
\bea\label{tophat}
\frac{\rho -\bar{\rho}}{\overline{\rho}} &=&
\begin{cases}
     \delta &,  0\le x <{\mathtt a},  \\
     -\delta/({\mathtt a}^{-3}-1) &,  {\mathtt a} \le  x < 1,
\end{cases}
\eea
where $\bar{\rho}$ is the cosmic mean at the lens redshift,
the parameter ${\mathtt a}$ delineates the over and under-dense regions, and $-1\le\delta\le ({\mathtt a}^{-3}-1)$ is the density contrast of the inner region.
When $\delta$ is negative this is a void model and when positive a  model for an overdensity.
The density contrast of the outer region $({\mathtt a}<x\le 1)$ is entirely determined by the necessity  of compensating for the excess/depleted central  density.
For this simple top-hat lens model we find
\be
\int_0^1{\frac{f(x',z_d)-f_{\rm RW}(x')}{x'}{dx'}}=-\delta\, \frac{\log {\mathtt a}}{({\mathtt a}^{-3}-1)},
\label{S0}\ee
and the structure term from  Eq.\,(\ref{S}) is
\be
{\cal S}(z_d)=-\frac{\delta \log {\mathtt a}}{({\mathtt a}^{-3}-1)} -(1+z_d)\frac{\log {\mathtt a}}{({\mathtt a}^{-3}-1)}\frac{d \delta}{dz_d}
-(1+z_d)\,\delta\,\frac{d \ \ }{d z_d}
\left[\frac{\log {\mathtt a}}{({\mathtt a}^{-3}-1)}\right].
\label{Szd}
\ee

\begin{figure}[!htbp]
\includegraphics[width=0.8\textwidth,height=0.4\textheight]{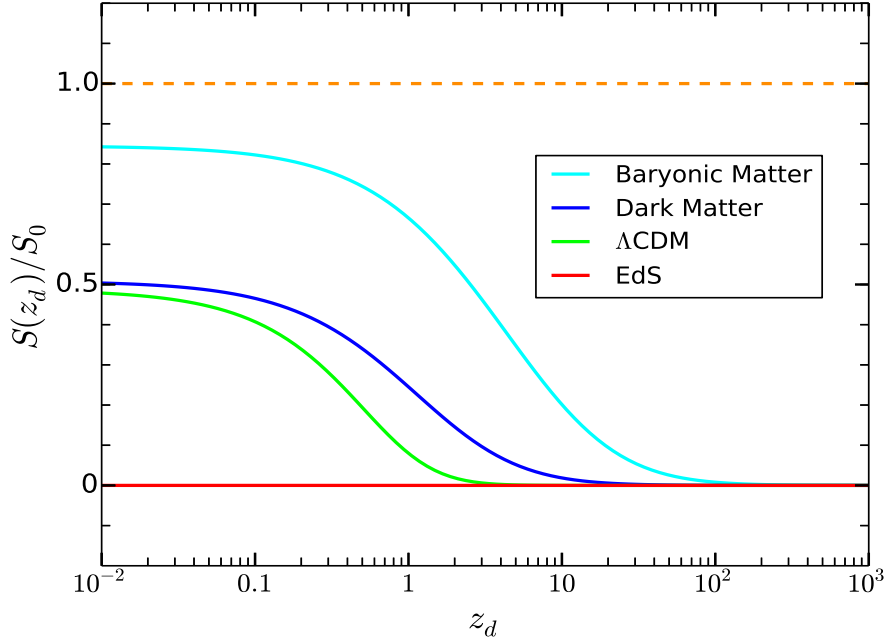}
\caption{Plots of the redshift evolution of the structure function ${\cal S}(z_d)/{\cal S}_0$ where ${\cal S}_0\equiv -\delta_0\log {\mathtt a}_0/ ({\mathtt a}_0^{-3}-1)$ for cluster ($S_0>0$) and void ($S_0<0$) models, see Eq.\,(\ref{tophat}).
The dashed brown curve is for cluster or void lenses co-expanding with the background cosmology.
The solid curves are computed assuming linear evolution in four background cosmologies (bottom to top): the Einstein de Sitter universe, $(\Omega_{\rm m},\Omega_\Lambda)=(1, 0)$, red; a $\Lambda$CDM universe, $(\Omega_{\rm m},\Omega_\Lambda)=(0.3,\,0.7)$, green; a dark matter only universe, $(\Omega_{\rm m},\Omega_\Lambda)=(0.3,\,0),$ blue; and a baryonic-matter only universe, $(\Omega_{\rm m},\Omega_\Lambda)=(0.05,\,0),$ cyan.
}
\label{fig:S}
\end{figure}

If the lens does not evolve in co-moving space, i.e., if  $\delta$ and ${\mathtt a}$ are both constants, the density perturbation is not evolving in size or shape relative to the background cosmology and $S(z_d)$ is just a constant given by Eq.\,(\ref{S0}).
We refer to this non-evolving, i.e., co-expanding, value as $S_{0}$, see the horizontal dashed brown curve in Fig.\,\ref{fig:S}.
If the density perturbation evolves relative to the FLRW background then $\delta$ and/or ${\mathtt a}$ are functions of the deflectors redshift $z_d$, the quantity given by Eq.\,(\ref{S0}) evolves with time, and the additional derivative terms in Eq.\,(\ref{Szd}) are present.
The structure term ${\cal S}(z_d)$ will depend on the background cosmological parameters if either of the two parameters $\delta$ or ${\mathtt a}$ does.
If $\delta$ evolves but ${\mathtt a}$ does not the second term is present and the perturbation's  amplitude evolves relative to the background cosmology but the perturbation  doesn't change its shape.
Linear perturbations are of this type (see the four cosmological parameter dependent curves in Fig.\,\ref{fig:S}).
If ${\mathtt a}$ evolves the last term is present and the shape of the perturbation evolves.
Relaxed clusters (see Fig.\,\ref{fig:Scluster}) and voids
produced by explosive motion are of this type \cite{Ostriker81,Bertschinger85a,Bertschinger85b}.
We next discuss linear perturbations and relaxed clusters in more detail.

\subsection{Linearly Evolving Cosmic Voids and Large Scale Overdensities}

As a first example of the ISW-z test using Eq.\,(\ref{newtest-thetaI}) we assume linear growth for the clusters or voids of the form given in Eq.\,(\ref{tophat}).
The fractional comoving radius of the top-hat remains constant  (${\mathtt a}={\mathtt a}_0$) while $\delta$ evolves as
\be
\delta=D_{+}(z_d)\,\delta_0,
\label{linear}
\ee
where the linear perturbation growth factor \cite{Heath77}
\be
D_{+}(z)=E(z)
\int_z^\infty \frac{(1+z^\prime)}{[E(z^\prime)]^3}\,dz^\prime
\Bigg/
\int_0^\infty \frac{(1+z^\prime)}{[E(z^\prime)]^3}\,dz^\prime,
\label{D+}
\ee
depends on the cosmological parameters through $E(z)$, see Eq.\,(\ref{E}).
Consequently, linear evolution  produces an evolving structure dependent term ${\cal S}(z_d)$ that depends on  cosmological parameters such as $\Omega_{\rm m}$ and $\Omega_\Lambda$,
\be\label{S_linear}
{\cal S}(z_d)=-\delta_0\, \frac{\log {\mathtt a}_0}{({\mathtt a}_0^{-3}-1)}\left[D_{+}(z_d)+(1+z_d)\frac{dD_{+}(z_d)}{d z_d}\right].
\ee

In Fig.\,\ref{fig:S} we have plotted  ${\cal S}(z_d)\div[-\delta_0\log {\mathtt a}_0/({\mathtt a}_0^{-3}-1)]$ for the top-hat cluster/void models of Eq.\,(\ref{tophat}) to illustrate evolution of the structure parts of Eqs.\,(\ref{newtest-rs}) and (\ref{newtest-thetaI}). To obtain the $z_d$ dependence of the structure part for a particular lens simply multiply each curve by the appropriate value of  $S_0\equiv[-\delta_0\log {\mathtt a}_0/({\mathtt a}_0^{-3}-1)]$.
To understand why the various evolutionary schemes produce different central temperatures at $z_d=0$ for exactly the same perturbation density at $z_d=0$, one has only to identify the two sources of the $z_d$ dependence in Eq.\,(\ref{S}).
When the derivative acts on the $(1+z_d)$ term the contribution to Eq.\,(\ref{calT}) is $H_dT_p/(1+z_d)$ which is directly proportional to the potential part of the lensing time delay $T_p.$
At the delayed exit, time the background CMB photons have further cooled and reddened whereas the lensed CMB photons, stuck in the lens, were not so reddened, and hence appear relatively bluer. This time-delay contribution to the temperature shift of the CMB is common to all evolutionary schemes and constitutes the entire temperature shift if the lens mass structure is evolving exactly like the background cosmology, i.e., co-expanding.
If the lens density evolves differently than the background, transiting CMB photons can loose or gain energy by virtue of the changing depth of the transited gravitational potential within the
lens.
If the lens is more condensed in the past, $\partial f(x,z_d)/\partial z_d>0$  in Eq.\,(\ref{S}), the fractional projected lens mass $f(x,z_d)$ decreased with  cosmic time and transiting photons lose  less energy when climbing out of the lens' potential well than they gain when falling in.
They would thus appear even bluer because of the evolution. However, if  $\partial f(x,z_d)/\partial z_d<0$, the lens structure is becoming more condensed with time (as shown in Fig.\,\ref{fig:S} for linear perturbations) and the CMB photons are redshifted  because of evolution.
The larger the evolution rate the more  reduction takes place in the time-delay blue shift.
In the EdS universe $D_+(z_d)\propto R(t_d)$ assuming linear evolution (see Eq.\,(\ref{S_linear})), and the evolution reddening completely cancels the time-delay blue shift (see the solid red curve in Fig.\,\ref{fig:S}).
See \citet{Chen13a} for more discussion about the time-delay and evolutionary contributions to the ISW effect.

\subsection{Virialized Clusters}
We evaluate the structure function $S(z_d)$ for four cluster models, one completely static, one co-expanding, and two virialized but accreting (see Table 1).
A virialized cluster model of the top-hat form given in Eq.\,(\ref{tophat}) that has stopped evolving, i.e., one that is completely static, would have a constant physical radius $r_{\mathtt a}$ and a constant central density $\rho_c$, or equivalently a $z$-dependent comoving radius
\be
{\mathtt a}=\frac{r_{\mathtt a}}{r_d(z)}={\mathtt a}_0(1+z),
\label{static-ra}
\ee
and a $z$-dependent central density enhancement
\be
1+\delta(z)\equiv\frac{\rho_c}{\overline{\rho}(z)}=(1+\delta_0)\left(1+z\right)^{-3}.
\label{static-rho}
\ee
The mass contained in the cluster is
\be
M_c(z)=\frac{4}{3}\,\pi\, r_{\mathtt a}^3\,\rho_c={\mathtt a}^3[1+\delta(z)] M_d,
\label{Mc}
\ee
where $M_d=4/3\,\pi\, r_d^3\,\overline{\rho}(z)$ is the (constant) mass contained in the co-moving Swiss cheese sphere from which the cluster condensed. The mass remaining in the depleted region around the cluster is simply $(M_d-M_c)$ and is the source of the accreting mass.
If the cluster wasn't virialized but simply co-expanding with the background cosmology, $\mathtt a$ and the central density enhancement $1+\delta$ would both remain constant.
For either of these two scenarios: model I. A static with constant physical radius $r_{\mathtt a}$ and constant central density $\rho_c$ or model II. A co-expanding cluster with constant fractional comoving radius $\mathtt a$ and constant central density contrast $\delta$, the mass of the central cluster remains constant.
Even though there are no observational measurements of mass accretion rates for clusters, they are all widely believed to be accreting mass.
To use the simple two-component mass profile models of the form given in Eq.\,(\ref{tophat}) to estimate the effect of accretion on the structure function  ${\cal S}(z_d)$  we  allow mass from the depleted background material surrounding the cluster $(r_{\mathtt a}\le r\le r_d)$, whose mass density is $[1-\delta/( {\mathtt a}^{-3}-1)]\times \overline{\rho}$, and who's partially depleted contents previously collapsed to form the cluster, to be continually falling onto that cluster after virialization at some specified rate
 \be
\frac{dM_c}{dt}=-(1+z)H_d\frac{dM_c}{dz}.
\label{Mdot}
\ee
If mass accretes, ${\mathtt a}$ and/or $\delta(z)$  differ from  the values given by Eqs.\,(\ref{static-ra}) and/or (\ref{static-rho}).
To make use of current accretion rate estimates we choose two accreting scenarios: the first, model III. Keep the central density  contrast fixed with $\rho_c(z)=(1+\delta)\,\overline{\rho}(z)=200\,\overline{\rho}(z)$ and let the comoving radius ${\mathtt a}(z)$ increase with cosmic time (decrease with redshift), see Eq.\,(\ref{Mc}).
In the second  accreting scenario, model IV. Assume the physical radius $r_a={\mathtt a}(z)r_d(z)$ remains constant while the central density enhancement $1+\delta(z)$  increases with time to accommodate the accreting mass. A comparison of these four simple cluster models is shown in Table~\ref{tab:cluster}.

Even though there is no observational data giving $dM_c/dz$ for accreting clusters, there are several fitting formulae arrived at by using the extended Press-Schechter formalism and N-body simulations to estimate that rate.
Two such simple fitting formulae are  2-parameter expressions due to McBride et al \citep{McBride09}
\be
M_c(z)/M_c(0)=(1+z)^\beta e^{-\gamma z},
\label{McBride}
\ee
and van den Bosch \cite{Bosch02}
\be
M_c(z)/M_c(0)=\exp\left\{\ln(1/2)\left[\frac{\ln(1+z)}{\ln(1+z_f)}\right]^\nu\right\}.
\label{Bosch}
\ee
These models are designed to represent stochastic averages of field galaxies and groups merging with the cluster and increasing its mass at the specified rate.
Other fitting formulas can be found in \cite{Bond91,Lacey93}.
Estimates of the parameters $(\beta,\gamma)$ and $(z_f,\nu)$ can be found in \cite{Boni15} for the  conventional $\Lambda$CDM background.

\begin{figure}[!htbp]
\includegraphics[width=0.8\textwidth,height=0.4\textheight]{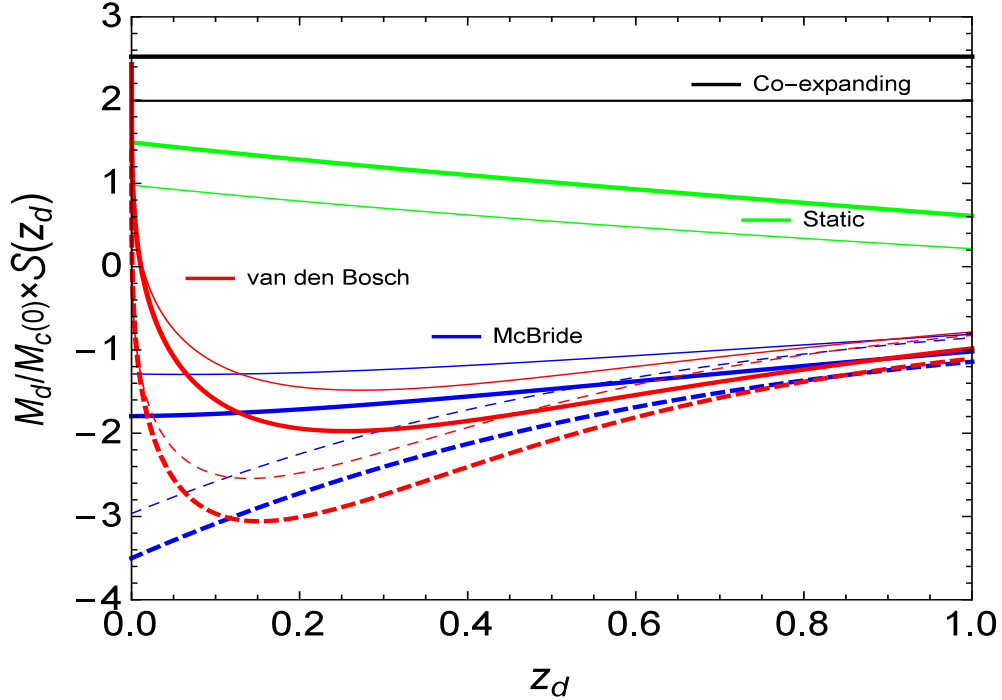}
\caption{
Structure function $S(z_d)$ for four cluster models from Table 1.
The effect of mass accretion on  $M_d/M_c(0)\times{\cal S}(z_d)$ in the flat $\Lambda$CDM background cosmology is plotted for clusters whose mass at $z=0$ is  $M_c(0)=10^{15}$ M$_\odot$. The $M_d/M_c(0)$ factor is included for the purpose of comparing $\Delta {\cal T}$'s of Eq.\,(\ref{newtest-rs}) because $r_{\rm s}\propto M_d$.
Thick lines are for clusters condensed from homogeneous spherical regions containing a total mass $M_d=10\,M_c(0)$, and thin lines for $M_d=2\,M_c(0)$.
Consequently clusters in larger lenses (thick curves) have smaller fractional radii $\mathtt a$  resulting in  different  $S(z_d)$ values, see Eq.\,(\ref{Szd}).
The two top horizontal black curves are for co-expanding non-accreting clusters and the next two straight green lines are for static non-accreting clusters.
The curved lines are accreting models. Blue models accrete according to  \cite{McBride09} (McBride) and red models according to  \cite{Bosch02} (van den Bosch). 
The solid line models assume the cluster's central density enhancement remains at $(1+\delta)=200$ during accretion and the physical radius $r_{\mathtt a}$ increases. The dashed line models assume the physical radii $r_{\mathtt a}$ remain constant and the central density enhancements  $1+\delta(z)$ increase with time.
Even though  accretion histories described by Eqs.\,(\ref{McBride}) and (\ref{Bosch}) are very similar, the associated accretion rates  are very different near $z\sim 0$ and  estimated values for ${\cal S}(z_d)$ based on these rates, clearly, cannot be trusted for $z_d<0.1.$
}
\label{fig:Scluster}
\end{figure}

In Fig.\,\ref{fig:Scluster} we estimate the effect of mass accretion,  Eq.(\ref{Mdot}), on the structure function ${\cal S}(z_d)$ of  clusters by assuming  mass accretion histories of the form  Eqs.(\ref{McBride}) and (\ref{Bosch}). Because of the limited availability of fitting data  we choose for all four cases shown, a cluster mass of  $M_c(0)=10^{15} M_\odot$ at $z=0$ and a central density enhancement $1+\delta_0=200$ at $z=0$ in a concordance background.
We take $(\beta,\gamma)=(-0.690, 1.280)$ and $(z_f,\nu)=(0.381, 1.252)$ in Eqs.\,(\ref{McBride}) and (\ref{Bosch}) respectively from fits found in \cite{Boni15}.

\begin{table}
\caption{\label{tab:cluster} Cluster models}
\begin{ruledtabular}
\begin{tabular}{cccccc}
Model & $M_c$ constant 	& $r_{\mathtt a}$ constant & ${\mathtt a}$ constant & $\delta$ constant & descriptions  \\
\hline
I  	&  yes 			&  yes				& 	no 			& no 			& non-accreting static  		\\
II 	&  yes 			&  no					&	 yes			& yes 		&  non-accreting co-expanding \\
III  	&  no 			&  no					& 	no 			& yes		& accreting with fixed $\delta$ \\
IV 	&  no 			&  yes				&	 no			& no 		&accreting with fixed $r_{\mathtt a}$\\
\end{tabular}
\end{ruledtabular}
\end{table}

The top four curves give ${\cal S}(z_d)$ of  Eq.\,(\ref{S}) for  non-accreting (constant cluster mass $M_c$) models I and II, the static and co-expanding models (black and green curves respectively).
The thick lines are cluster models whose cluster mass $M_c(z)$ at $z=0$ is 1/10 of the Swiss cheese void's constant mass $M_d$ (the mass of the homogeneous sphere from which the cluster condensed).
The thin lines are for smaller void masses, $M_d=2 M_c(0)$ with correspondingly smaller radii $\propto (2/10)^{1/3}$.

The blue and red curves in Fig.\,\ref{fig:Scluster} are computed assuming the McBride and van den Bosch accretion histories of Eqs.\,(\ref{McBride}) and (\ref{Bosch}) respectively.
The solid and dashed curves are for constant central density contrast $\delta$ (model III) and constant physical radius $r_{\mathtt a}$ (model IV) respectively, see Table 1.
For $z\le 0.1$ estimated  values of ${\cal S}(z_d)$ cannot be accurate because the fitted values of $dM(z)/dz$ differ significantly between Eqs.\,(\ref{McBride}) and (\ref{Bosch}).
In general the mass accreted while the CMB photons transit the cluster  deepens the potential well from which the photons must climb to reenter the background cosmology.
They are thus reddened by accretion.
As can be seen in Fig.\,\ref{fig:Scluster} the effect of accretion seems to dominate the ISW  effect for cluster centers.

The large negative values of ${\cal S}(z_d)$ seen in Fig.\,\ref{fig:Scluster} for the two accreting cluster models III and IV are highly uncertain but are clearly worrisome since they would make the CMB temperature in the center of the cluster much cooler than expected, and would even produce a central cold spot.
This is in contrast to conventional wisdom that galaxy clusters produce CMB hot spots at $z\lesssim1$ because  the accelerated expansion of the Universe reduces the depth of the potential well from which the transiting photons must climb.
We find that when accretion is happening its effects can overwhelm the effects of acceleration, i.e., the potential well might be deepening inspite of the acceleration.
Of course the accuracy of the predictions made by these models can be questioned because they assume continuous accretion whereas the consensus is that mass accretes via discrete mergers of halos associated with galaxies groups ($\le 10^{13} M_\odot$) and/or  galaxies ($\le 10^{12} M_\odot$). To estimate the reasonableness  of the continuous accretion model we compare the photon's lensing time with the time between mergers. The time it takes a photon to cross the Swiss cheese void is $\sim 2.5\times 10^{-1}$ Gyr for a $10^{15}M_\odot$ cluster lens with $M_d/M_c(0)=10$ and  $\sim 1.5\times 10^{-1}$ Gyr when  $M_d/M_c(0)=2$. Time between mergers is $\sim 7\times 10^{-3}$ Gyr for galaxies and $\sim 7\times 10^{-2}$ Gyr for groups. If only galaxies were accreting then $\sim 35$ (21 for $M_d/M_c(0)=2$) would merge while the photon transited the lens whereas if groups were responsible for the mass accretion only $\sim 3.5$ (2.1 for $M_d/M_c(0)=2$) would have merged.
The time taken for photons to cross the central cluster itself is only $\sim 2.0\times 10^{-2}$ Gyr; however, it is the change of the potential across the entire compensated lens that determines the ISW temperature shift, i.e., transiting CMB photons have to make their way across the large continually depleting low density compensating region before reentering the background cosmology. We consider two extreme physical pictures in the following. A more realistic picture would be a combination of the two. If $ 10^{15} M_\odot$ clusters assemble their masses by accreting only  $10^{12} M_\odot$ galaxies there would be $\sim 1000$ galaxies in the cluster and $\sim 9000$ galaxies surrounding the cluster in the low density compensating region (for the $M_d=10 M_c(0)$ Swiss cheese lens). During the time it took the CMB photons to cross the entire lens $\sim 35$ galaxies would have accreted from the compensating region to the central cluster. If clusters assembled their masses by accreting $10^{13} M_\odot$ groups, all numbers would be decreased by a factor of 10. If the low density compensating region surrounding the cluster was smaller, i.e., if $M_d=2 M_c(0)$ then there would be only 1000 galaxies in the compensating region and during the CMB crossing only $\sim 21$ galaxies would have accreted.
The current state of accretion theory, as judged by Fig.\,\ref{fig:Scluster}, clearly suggests that Eq.\,(\ref{newtest-rs}) is more likely to constrain cluster structure and evolution than cosmology at this time. If accurate density profiles can be constructed, combined with additional measurements of central cluster temperatures deficits one should be able to put limits on cluster mass accretion rates.

\section{Conclusions}

The ISW effect has been recently detected via the aperture photometry method (stacking/averaging patches of the CMB maps around known cosmic voids or galaxy clusters) by several groups \cite{Granett08,Hernandez10,Planck14,Ilic13,Cai14} and future observations promise more and better data.
We present a new method of using this data to potentially constrain the cosmological parameters by applying the ISW effect to individual inhomogeneities such as galaxy clusters and cosmic voids. We were able to develop this ISW-redshift test
only after discovering a simple relation between the Fermat potential of an embedded lens and the frequency shift of photon crossing that lens.
However, to use this test  to extract the Hubble parameter and/or the curvature parameters the evolution of the lens has to be well understood.
We have illustrated use of the ISW-z test by constructing models for clusters and voids with very simple density profiles and simple evolutions (i.e., top-hats for linearly evolving clusters/voids and completely virialized clusters with and without accretion).
However, for both galaxy clusters and cosmic voids, neither their density profiles nor their time evolution is currently well enough constrained by observations to be used in this test.
Consequently, the proposed  ISW-z test  might be more appropriately used to measure structure functions ${\cal S}(z_d)$, and constrain dark matter profiles, evolution, and accretion by assuming a specific cosmology  and using the CMB observations.
There are several theoretical/numerical papers modeling the formation and evolution of cosmic voids \cite{Ostriker81,Bertschinger85a,Bertschinger85b,Sheth04} which can be used to estimate the structure term ${\cal S}(\theta_I,z_d)$ of Eq.\,(\ref{S}) and the ISW-z test can possibly confirm or reject such models as more data becomes available.
The density profile of galaxy clusters is thought to be much better constrained than profiles of cosmic voids, and for their low redshift evolution it is reasonable to assume that they are virialized.
However, as seen in Fig.\,\ref{fig:Scluster} the structure term ${\cal S}(\theta_I,z_d)$ is sensitive to the accretion rate which is  poorly understood. What is clear from Fig.\,\ref{fig:Scluster} is that accretion can make the centers of clusters appear unexpectedly cool just as is currently being seen \cite{Planck15}.
Obviously better models for cosmic voids and galaxy clusters are needed before the ISW-z test will constrain the cosmological parameters.

\label{lastpage}

\end{document}